*Fuad Aleskerov, Natalia Meshcheryakova,*
*Sergey Shvydun*

# CENTRALITY MEASURES IN NETWORKS BASED ON NODES ATTRIBUTES, LONG-RANGE INTERACTIONS AND GROUP INFLUENCE

We propose a new method for assessing agents’ influence in network structures, which takes into consideration nodes attributes, individual and group influences of nodes, and the intensity of interactions. This approach helps us to identify both explicit and hidden central elements which cannot be detected by classical centrality measures or other indices.

*Aleskerov Fuad*, National Research University Higher School of Economics (HSE), International Laboratory of Decision Choice and Analysis, V.A. Trapeznikov Institute of Control Sciences of Russian Academy of Sciences (ICS RAS), Moscow; alesk@hse.ru

*Meshcheryakova Natalia*, National Research University Higher School of Economics (HSE), International Laboratory of Decision Choice and Analysis, V.A. Trapeznikov Institute of Control Sciences of Russian Academy of Sciences (ICS RAS), Moscow; natamesc@gmail.com

*Shvydun Sergey*, National Research University Higher School of Economics (HSE), International Laboratory of Decision Choice and Analysis, V.A. Trapeznikov Institute of Control Sciences of Russian Academy of Sciences (ICS RAS), Moscow; shvydun@hse.ru

# 1. Introduction

During last years there has been a deep interest in the analysis of different communities and complex networks, specially their structure and key elements detection. Most classical measures do not take into account individual properties of each element. Additionally, they do not completely take into account the intensities of interactions between elements, especially, long-range interactions. One more problem arises from the fact that not only one node but also a group of nodes can influence other nodes. Consequently, the results of the application of classical measures inadequately represent the actual state of a system.

Existing measures are not accurate even for small networks. There exist several simple network structures where classical indices do not elucidate hidden elements influential in the network. This can be explained by the fact that these indices do not fully take into account individual properties of nodes, the intensity level of direct connections and long-range interactions between nodes of the networks. For instance, classical centrality measures do not pay attention to the possibility of chain reactions of a system (so-called domino or contagion effect). The incessant changes in composition and structure of groups and nets magnify the complexity of the problem.

The main objective of our research is to develop new efficient methods of key nodes detection which take into account these particular aspects of the problem under consideration.

The paper is organized as follows. In Section 2 we provide a review of existing methods of key nodes detection in networks and demonstrate some of their shortages. In Section 3 we formally describe the new method and show how it works on a simple example. We also emphasize advantages and weaknesses of the proposed method.

**Acknowledgements**

The paper was prepared within the framework of the Basic Research Program at the National Research University Higher School of Economics (HSE) and supported within the framework of a subsidy by the Russian Academic Excellence Project "5–100". The work was conducted by the International Laboratory of Decision Choice and Analysis (DeCAn Lab) of the National Research University Higher School of Economics.

## 2. Literature review

There have been developed many indices to measure the centrality level of each node. Some of them are based on the number of links to other nodes. Other techniques consider how close each node is located to other nodes of the network in terms of the distance, or how many times it is on the shortest paths connecting any given node-pairs. There are also some indices based on ideas from cooperative game theory and voting theory. These indices are called centrality measures.

Consider network-graph $G = \{V, E, W\}$, where $V = \{1, \dots, n\}$ is the set of nodes, $|V| = N$, $E \subseteq V \times V$ is the set of edges, and $W = \{w_{ij}\}$ is the set of weights – real numbers prescribed to each edge $(i, j) \in E$. Network-graph $G$ is directed if $\exists i, j \in V: (i, j) \in E \;\&\; (j, i) \notin E$ and is undirected otherwise. The graph is called unweighted if $\forall i_1, i_2, j_1, j_2 \in V: (i_1, j_1) \in E \;\&\; (i_2, j_2) \in E \Rightarrow w_{i_1 j_1} = w_{i_2 j_2}$, i.e. every edge has the same weight. Below we consider only directed weighted graphs, i.e., the set of pairs $(i, j) \in E$ is ordered.

A network-graph $G$ can also be represented in the form of matrix $A = [a_{ij}]_{N \times N}$, where $a_{ij} = 1$ if $(i, j) \in E$ and $a_{ij} = 0$ otherwise, or in the form of matrix $W = [w_{ij}]_{N \times N}$, where $w_{ij}$ is a weight that indicates the intensity of connection of node $i$ to node $j$. The matrix $A$ is called an adjacency matrix of the network-graph $G$ while the matrix $W$ is called a weighted adjacency matrix of the network-graph $G$. In terms of influence, $a_{ij} = 1$ means that node $i$ influences node $j$; for weighted graphs, if $w_{ij} > 0$ then node $i$ influences node $j$ with power $w_{ij}$, otherwise, node $i$ does not influence node $j$ ($a_{ij} = 0$ or $w_{ij} = 0$). Additionally, the nodes can also have individual attributes (for instance, weights) that will be denoted by $u_i^k$, where $i$ is a node number and $k$ is the number of the attribute, $k \in K$.

Denote by $\vec{N}_i = \{j \in V: (i, j) \in E\}$ a set of neighbors of node $i$ which $i$ is connected to, $\overleftarrow{N}_i = \{j \in V: (j, i) \in E\}$ is a set of neighbors of node $i$ that are connected to $i$, $\overline{N}_i = \vec{N}_i + \overleftarrow{N}_i = \{j \in V: (i, j) \in E \text{ or } (j, i) \in E\}$ is a set of all neighbors of node $i$ in a network-graph $G$.

### *2.1. Degree centralities*

The simplest centrality measure is the degree centrality that is calculated for undirected network-graphs as the total number $C_i^{deg}$ of $i$'s neighbors for each node $i$ [Freeman, 1979]:

$$C_i^{deg} = |\bar{N}_i|.$$

High values of the degree centrality identify nodes with the highest number of connections to other nodes, i.e. nodes for which it is easier to gain access to and/or influence over other nodes. A central node occupies a structural position (network location) that serves as a source or conduit for larger volumes of information exchange or other resource transactions with other nodes.

For directed network-graphs four versions of degree centrality measure are possible

- In-degree centrality – the number of in-coming edges to a node

$$C_i^{in-deg} = |\overleftarrow{N}_i|.$$

High values of in-degree centrality mean that a node is strongly affected by its neighbors. Alternatively, low values of in-degree centrality identify nodes that are not influenced by other nodes.

- Out-degree centrality – the number of out-going edges from a node

$$C_i^{out-deg} = |\overrightarrow{N}_i|.$$

High values of out-degree centrality represent the influence power of a node, i.e. the higher the value the more nodes are under its control. Conversely, low values of out-degree centrality mean that a node has a small effect on its neighbors.

- Degree centrality – the total number of $i$'s neighbors

$$C_i^{total\ deg} = C_i^{in-deg} + C_i^{out-deg}.$$

This measure is obtained by ignoring directions of edges and high values of total degree centrality identify the most active nodes.

- Degree difference centrality – the difference between the number of out-going edges from a node and the number of in-coming edges to a node

$$C_i^{deg\ diff} = C_i^{out-deg} - C_i^{in-deg}.$$

In power networks high values of degree difference show the relative influence of a node on its neighbors.

For weighted degree network-graphs it is also possible to calculate the degree centrality with respect to the weights of adjacent edges. Then four measures are introduced

- Weighted in-degree centrality

$$C_i^{w\ in-deg} = \sum_{j \in V:\,(j,i) \in E} w_{ji} = \sum_{j=1}^{n} w_{ji}.$$

- Weighted out-degree centrality

$$C_i^{w\ out-deg} = \sum_{j \in V:\,(i,j) \in E} w_{ij} = \sum_{j=1}^{n} w_{ij}.$$

- Weighted degree centrality

$$C_i^{total\ w\ deg} = C_i^{w\ in-deg} + C_i^{w\ out-deg}.$$

- Degree difference centrality

$$C_i^{w\ deg\ diff} = C_i^{w\ out-deg} - C_i^{w\ in-deg}.$$

The interpretation of weighted degree centralities is practically the same as for unweighted degree centralities but weighted measures are more representative than unweighted ones due to the fact that weighted networks consider the intensities of connections.

Since the degree centrality measures do not consider the strength of adjacent nodes, i.e., information about the degree centrality of adjacent nodes, there have been developed several indices which take into account this feature. A generalization is what is known as an eigenvector centrality that considers not only neighboring but also long-distance connections. Basically, this measure is applicable to symmetric relations. It assigns relative scores to all nodes in a network based on the concept that connections to high-scoring nodes contribute more to the score of the node in question than equal connections to low-scoring nodes. If we talk about asymmetric relations as networks of influence it is more valuable to influence powerful nodes.

The calculation of the centrality measure for each node is related to an eigenvalue problem with respect to weighted adjacency matrix $W$ of a network-graph: a vector of relative centrality $C^{eigen}$ is an eigenvector of the adjacency matrix, i.e.

$$W \cdot C^{eigen} = \lambda \cdot C^{eigen}.$$

Generally, all eigenvectors of the matrix $W$ can be considered as a centrality measure. However, an eigenvector that corresponds to a maximal eigenvalue is more preferable: by Perron-Frobenious theorem this vector (and only this except its co-directional vectors) is positive and real for irreducible non-negative matrix $W$ [Gantmacher, 2000], i.e., for a graph which is strongly connected.

This approach to centrality evaluation was proposed by P. Bonacich [Bonacich, 1972] and is sometimes known as Bonacich's index. [Bonacich, 1987] considers a generalization of this approach where a degree of nodes counted towards the centrality evaluation. As for an eigenvector centrality this measure is more representative for symmetric relation. For asymmetric graphs of influence the calculation is the same. Namely, a parametric family of centrality measures can be represented as

$$C_i^{Bonacich}(\alpha,\beta) = \sum_j \left(\alpha + \beta \cdot C_j^{Bonacich}(\alpha,\beta)\right) \cdot W_{ij}$$

or in a matrix form

$$C^{Bonacich}(\alpha,\beta) = \alpha \cdot (I - \beta \cdot W)^{-1} \cdot W \cdot 1,$$

where $I$ is an identity matrix and 1 is the unit vector.

Apparently, parameter $\alpha$ affects only the variance of a centrality vector. Parameter $\beta$ represents the degree to which a centrality of one node is a function of centralities of adjacent nodes. If a centrality of one node is a positive function of its neighbors' centralities then we select positive parameter $\beta$.

The main innovation is that this approach also considers negative values of parameter $\beta$. This leads to the fact that centralities of neighbors are negatively counted in node centrality, i.e. it is not beneficial to be connected with central nodes. Negative $\beta$ is usually required in bargaining networks where it is more profitable to be connected with weak players because powerful players have more potential trading partners, which reduces your bargaining power.

In practice, an eigenvector centrality is not very feasible especially for large networks because it gives a lot of zero centralities if there are many sparse cohesive components in a graph. There have been introduced (or used previously entered) other measures to overcome this shortage. Katz centrali-ty is one of such measures introduced in [Katz, 1953]. This centrality is defined as the solution of the two-parameter equation

$$C_i^{Katz}(\alpha,\beta) = \alpha \cdot \sum_j C_j^{Katz}(\alpha,\beta) \cdot W_{ij} + \beta$$

or in a matrix form

$$C^{Katz}(\alpha,\beta) = \beta \cdot (I - \alpha \cdot W)^{-1} \cdot 1,$$

where 1 is the unit vector.

The introduction of parameter $\beta$, which corresponds to the initial value of centralities, precludes the possibility of solution with zero components. In practice, parameter $\alpha$ is selected so that $\alpha < \frac{1}{\lambda_{max}}$, where $\lambda_{max}$ is the largest eigenvalue of the matrix $W$.

Katz centrality, in its turn, is not free from an essential fault: for a node with a high degree centrality value and a high Katz centrality value its neighboring nodes will also have high Katz centrality values even if their degree centrality values are not very high.

Some modifications of Katz centrality are used to overcome this disadvantage. For example, the PageRank centrality was proposed where degrees of adjacent nodes are introduced

$$C_i^{PageRank} = \alpha \cdot \sum_j \frac{C_j^{PageRank}}{C_j^{w\,out-deg}} \cdot W_{ij} + \beta$$

or in a matrix form

$$C^{PageRank} = \beta \cdot [I - \alpha \cdot W \cdot (C^{w\,out-deg})^{-1}]^{-1} \cdot 1,$$

where $C^{w\,out-deg} = diag\left(C_1^{w\,out-deg}, \dots, C_n^{w\,out-deg}\right)$ (if $C_j^{w\,out-deg} = 0$ then the corresponding summand is set to zero), 1 is the unit vector. This formula was taken as a basis in Google to rank search engine queries [Brin, Page, 1998].

### *2.2. Closeness centralities*

Besides the degree centralities, there are also methods that consider how close each node is located to other nodes of a network in terms of a distance. These measures indicate the level of closeness of each node and are called closeness centrality indices.

The standard closeness centrality measure for each node is equal to the value that is proportional to the harmonic mean of the length of the shortest paths between the $i$-th node and the rest of it in a network [Rochat, 2009]

$$C_i^{cl} = \sum_j \frac{1}{d_{ij}}.$$

### *2.3. Betweenness centralities*

There are also indices that show how many times a node is on the shortest paths connecting any given pair of nodes. These measures were proposed in [Freeman, 1977; Freeman et al., 1991; Newman, 2005] and are called the betweenness centrality measures. Versions for such centralities are

- the number of shortest paths passing through a given node

$$C_i^{btw} = \sum_{jk} \sigma_{jk}(i),$$

where $\sigma_{jk}(i)$ is the number of shortest paths that connect $j$ and $k$ and contain $i$;

- the relative number of the shortest paths passing through a given node and connecting two nodes to the total number of shortest paths connecting these nodes

$$C_i^{relative\ btw} = \sum_{jk} \frac{\sigma_{jk}(i)}{\sigma_{jk}},$$

where $\sigma_{jk}$ is the number of shortest paths that connect $j$ and $k$;

- the sum (throughout all pairs of nodes) of maximum flows from the first node of pair to the second one passing through a given node [Freeman et al., 1991]:

$$C_i^{flow\ btw} = \sum_{jk} m_{jk}(i),$$

where $m_{jk}(i)$ is a maximum flow from $j$ to $k$ that passes through $i$;

- the sum (throughout all pairs of nodes) of the mathematical expectations of the number of random walks connecting a pair of nodes and passing through a given node [Newman, 2005].

### *2.4. Centralities from cooperative game theory*

Many attempts of key nodes detection in networks came from cooperative game theory. In that case, a network is interpreted as a set of interacting individuals that contribute to a total productive value of a network and the problem is how to share generated value among them. In [Myerson, 1977] there was proposed a measure which is based on the power index and is a version of the Shapley-Shubik index [Shapley, Shubik, 1954] for communication games. The Myerson value has an allocation rule in the context of network games where the value of each individual depends on the value generated by a network with and without that individual. More precisely, the Myerson value is an average contribution of a node to all subgraphs of a graph with respect to some predefined values of subgraphs, i.e.

$$C_i^{MV}(G,v) = \sum_{S \in V} \frac{(|S|-1)!(|N|-|S|)!}{|N|!} (v(S) - v(S \backslash \{i\})),$$

where $S$ is a subgraph of graph $G$, $v(S)$ is some predefined value of subgraph $S$ and $v(S \backslash \{i\})$ is a predefined value of subgraph $S$ without node $i$.

### *2.5. Centralities from voting theory*

Existing measures are not accurate even on small networks. There exist several simple network structures where classical indices do not elucidate hidden elements influential in the network. This can be explained by the fact that these indices do not fully take into account individual properties of nodes, the intensity level of direct connections and interactions between nodes of the networks.

In [Aleskerov et al., 2014] a novel method for estimating the intensities of nodes' interactions was proposed. This method is based on the power index analysis that was worked out in [Aleskerov, 2006] to find the most pivotal agents in Russian Parliament (1999–2003) and adjusted for the network theory. The index (originally called a key borrower index) is a Short-Range Interaction Centrality (SRIC) that was employed to find the most pivotal borrower in a loan market in order to take into account some specific characteristics of financial interactions. An important feature of SRIC index is that it does not take into account all edges in a graph which is logical for many cases in networks. The choice of edges that are influential in a network depends on additional parameter $q_i$ which varies with the node $i$ and represents some critical threshold value.

The SRIC index is calculated for each node individually in order to determine the influence of other nodes to it. In that case, only direct neighbors are considered to estimate the direct and indirect influence to him/her. The intensity of direct influence $p_j^i$ of node $j$ to node $i$ is calculated as

$$p_j^i = \frac{w_{ji}}{\sum_k w_{ki}},$$

where $w_{ki}$ is a weight of an edge from node $k$ to node $i$, while the intensity of indirect influence $p_{jy}^i$ of node $j$ to node $i$ through node $y$ is calculated as

$$p_{jy}^i = \begin{cases} \frac{w_{jy}}{\sum_k w_{ki}}, if\ w_{yi} > 0, w_{jy} < w_{yi} \text{ and } y \neq j, \\ \frac{w_{yi}}{\sum_k w_{ki}}, if\ w_{yi} > 0, w_{jy} > w_{yi} \text{ and } y \neq j, \\ 0, otherwise. \end{cases}$$

After the intensity of influence to node $i$ of its adjacent nodes is calculated, a set of all possible critical groups of nodes for node $i$ is constructed. A group of nodes is critical if the total weight of edges from these nodes to the node $i$ is more than or equal to some pre-defined threshold $q_i$. The critical group is interpreted as a group that may influence a particular node.

After a set of critical groups for node $i$ is defined, we can identify a total number of groups where each node $j$ plays a pivotal role. A node $j$ is pivotal in a critical group if its exclusion from this critical group makes the group non-critical. The value of the index for each node reflects the magnitude of its pivotal role in the group. The higher the value, the more pivotal the node is. The most pivotal node will be the one that becomes pivotal in more critical groups than any other node does.

The total intensity of influence of node $j$ to node $i$ is aggregated over the intensities of all groups where the node $j$ is pivotal with respect to the size of the group. The influence of each node to node $i$ is equal to the normalized value of the final intensity measure.

After the total intensity of connection between node $i$ and its adjacent nodes is calculated, the index is aggregated over all nodes taking into account individual attributes of each node.

Unfortunately, the SRIC index also has some shortages. In the SRIC index only direct interactions of the first level are taken into account, which is not correct in some cases when long-range interactions play a pivotal role or where chain reactions are possible. Also, the SRIC index does not elucidate nodes that have a weak direct influence to particular node $i$ but are highly influential to its adjacent nodes (see Fig. 1). This is due to the fact that long-range interactions are not taken into account.

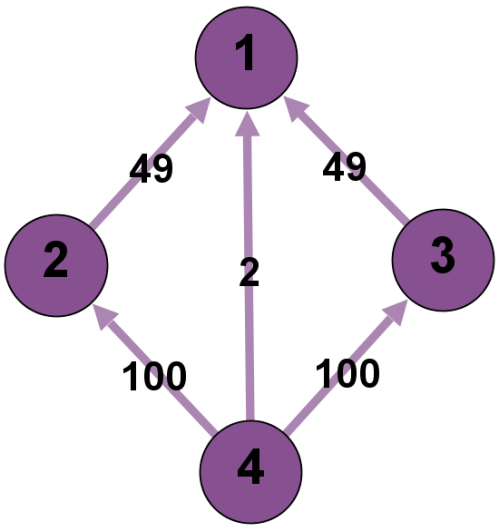

**Fig. 1.** SRIC: node 4 does not influence node 1 ($\forall i \in \{1,2,3,4\}\ q_i = 40$)

To demonstrate the shortages of existing measures consider the following Numerical Example 1 (see Fig. 2). There are 8 nodes in network-graph $G$ and the weights of edges are given in Fig. 2.

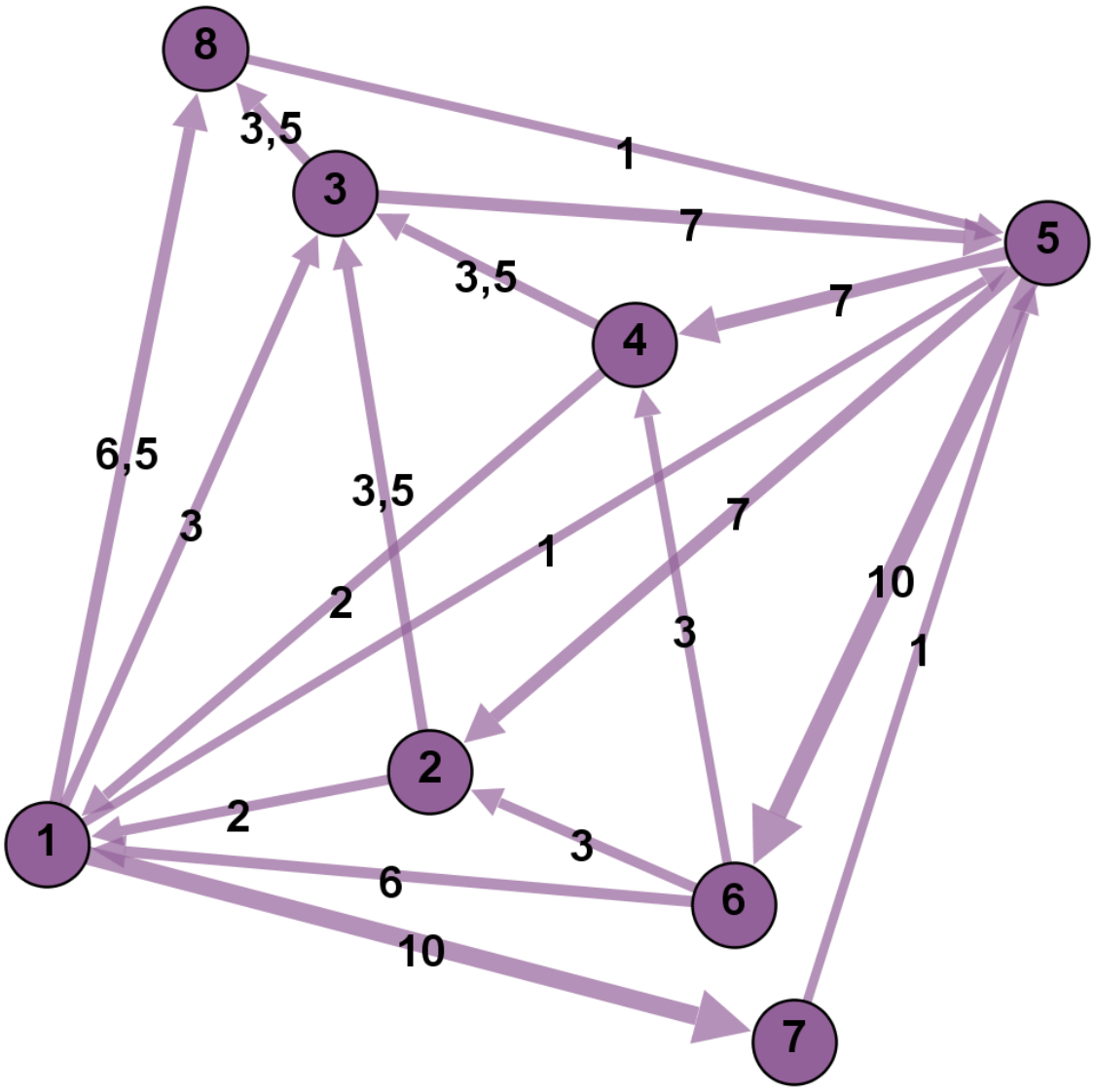

**Fig. 2.** Numerical Example 1

Let us evaluate classical centrality measures for Numerical Example 1 (Table 1).

**Table 1.** Classical centrality measures for Numerical Example 1

| Centrality / Node | $C^{w\ in-deg}$ | $C^{w\ out-deg}$ | $C^{w\ deg}$ | $C^{w\ deg-diff}$ | $C^{btw}$ | $C^{cl}$ (*) | $C^{PageRank}$ | $C^{eig}$ | $C^{Bonacich}$ |
|---|---|---|---|---|---|---|---|---|---|
| 1 | 10 | **20,5** | **30,5** | **10,5** | **13** | 0,0160 | 0,128 | 0,725 | –1,217 |
| 2 | 10 | 5,5 | 15,5 | –4,5 | 1,5 | 0,0126 | 0,103 | 0,589 | –0,890 |
| 3 | 10 | 10,5 | 20,5 | 0,5 | 6 | 0,0187 | **0,146** | 0,653 | –0,835 |
| 4 | 10 | 5,5 | 15,5 | –4,5 | 1,5 | 0,0126 | 0,103 | 0,589 | –0,890 |
| 5 | 10 | **24** | **34** | **14** | **24** | 0,0227 | **0,248** | **1** | –1,290 |
| 6 | 10 | 12 | 22 | 2 | 9 | 0,0155 | 0,107 | **0,802** | –1,253 |
| 7 | 10 | 1 | 11 | –9 | 0 | **0,0093** | 0,072 | 0,370 | **–0,708** |
| 8 | 10 | 1 | 11 | –9 | 0 | **0,0097** | 0,094 | 0,359 | **–0,708** |

(*) Closeness centrality: inverse average maximal outflow ⇒ low values are more significant.

According to weighted out-degree, weighted degree, degree difference and betweenness centrality measures (where high weights are better) nodes 1 and 5 are the most powerful in the network. Closeness centrality measure (where small weights are better) considers nodes 7 and 8 as the most powerful. If we take into account the strength of the neighbors, then nodes 3 and 5 will be chosen by PageRank, nodes 5 and 6 by eigenvector and, finally, nodes 7 and 8 by Bonacich centrality. Overall, we can conclude that nodes 1 and 5 are chosen by the most of centrality measures.

However, in most situations not all edges should be taken into account. Suppose now threshold level $q_i$ is 70% for each node $i$, i.e. node $i$ is influenced by individual node or a group of them only if their total influence to $i$ is more than or is equal to 70% of the total influence to $i$. Such information is not taken into account by classical centrality measures contrary to SRIC index.

In Fig. 3 we demonstrate substantial influence in the network for our Numerical Example 1.

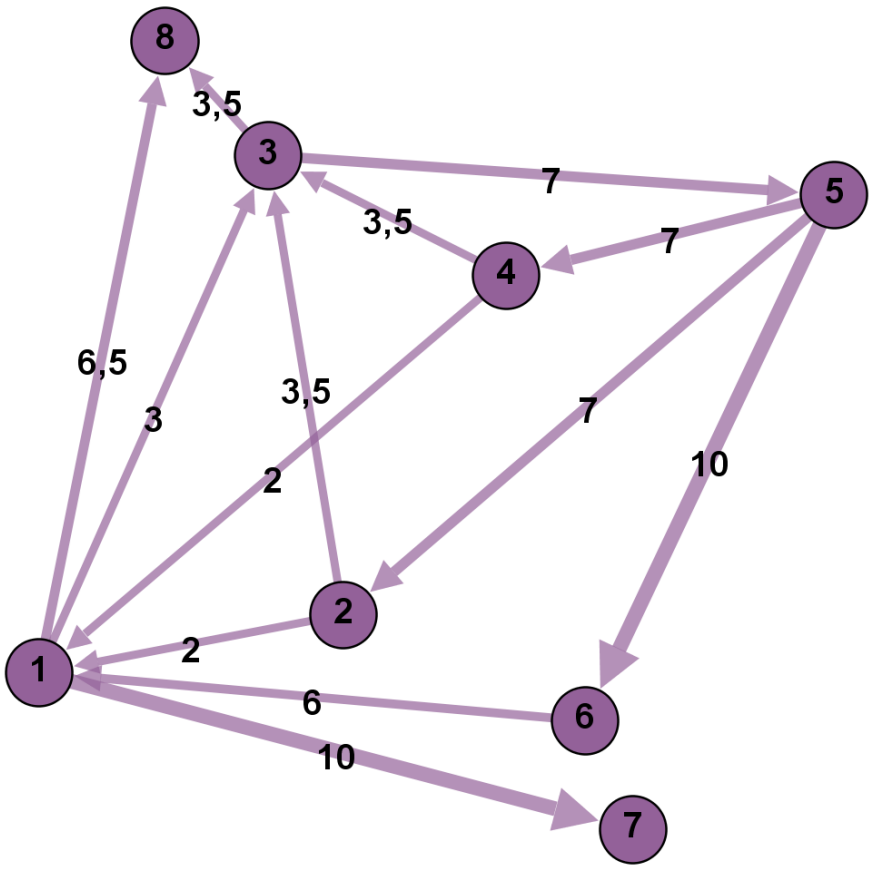

**Fig. 3.** Substantial influence for Numerical Example 1 ($q_i = 70\%$)

It should be mentioned that nodes 7 and 8 have no real influence on other nodes in the network. Such information is not taken into account by centrality measures, so the role of nodes 7 and 8 are overestimated.

The results of SRIC index are provided in Table 2.

**Table 2.** SRIC index for Numerical Example 1

| **Node** | 1 | 2 | 3 | 4 | 5 | 6 | 7 | 8 |
|---|---|---|---|---|---|---|---|---|
| **SRIC** | **0,216** | 0,072 | 0,159 | 0,072 | **0,375** | 0,106 | 0 | 0 |

SRIC index also identifies nodes 1 and 5 as the most influential in this network. The key improvement of SRIC index comparing to classical centrality measures is that SRIC index ignores insignificant connections (with respect to pre-defined thresholds $q_i$) and considers short connections (of length 1 and 2). Generally, neither classical centrality measures nor SRIC index consider long connections as well as group influence and individual attributes (as pre-defined thresholds), which leads to the fact that existing methods may not detect hidden influential nodes. Hence, all these methods underestimate the role of node 3 which in turn controls node 5 and also influences node 8. Due to the fact that node 3 significantly influences node 5, and node 5 is central in this graph we suppose that node 3 is more influential than node 5 (node 3 in a wild card).

## 3. Long-Range Interactions Centrality (LRIC)

We propose a new method for assessing the nodes influence in the network. Contrary to SRIC index, our methodology allows to consider interactions between nodes not just on the first level, but also on some levels beyond.

There are two different ideas on how to take into account long-range interactions between nodes of the network. The first one is a distance-based approach where all different paths are considered for each node and somehow aggregated into a single value. The second one is based on the idea of simulations where we analyze the influence of individual nodes and their combinations to a whole network. Both ideas have simple interpretations and can be applied to different networks.

The formulation of a problem is as follows: consider network-graph $G(V, E)$, where $V = \{1, \dots, n\}$ is a set of nodes, $|V| = N$, $E = \{(i, j),\ i, j \in V\}$ is a set of weighted edges, and $w_{ij}$ is a weight of edge $(i, j)$. The issue is to define the most influential nodes in this graph.

Let us consider the following graph where $N = 10$ (Fig. 4).

We propose two approaches to find central nodes in a network. This concept is motivated by the fact that indirect connections can play a significant role in different situations; however, classical centrality measures do not consider long interactions. For that reason we develop indices that take into account distant nodes. Generally, highly distant nodes do not influence other nodes of a graph; hence, we introduce a parameter $s$ that defines the lengths of connections we take into account. Accordingly, if long interactions do not influence indirect nodes then parameter $s$ is equal to 1, and contrary, if all levels of indirect connections matter, then parameter $s$ is unlimited.

Primarily, we introduce some basic definitions.

Let $\overleftarrow{N_i}$ be a set of directly connected nodes of node $i$ (incoming neighbors), i.e. $\overleftarrow{N_i} = \{j \in V | w_{ji} \neq 0\}$. Let every node has an individual attri-

bute – predefined threshold $q_i$, i.e. the threshold level when a node becomes affected.

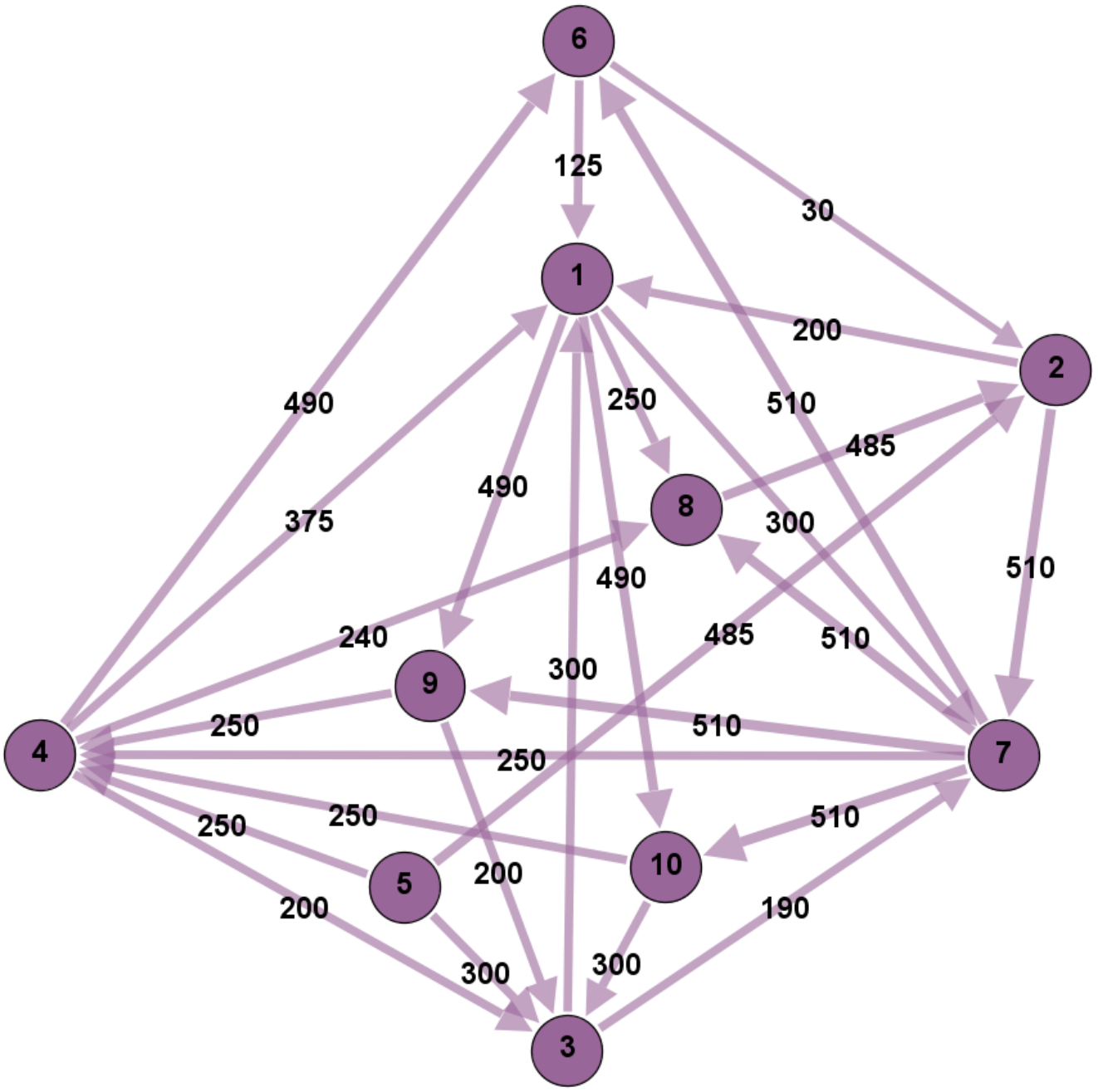

**Fig. 4.** Numerical Example 2

*Definition 1.* A group of neighbors of node $i$ $\Omega(i) \subseteq \overleftarrow{N}_i$ is critical if $\sum_{j \in \Omega(i)} w_{ji} \geq q_i$.

*Definition 2.* Node $k \in \Omega(i)$ is pivotal if $\sum_{j \in \Omega(i) \setminus \{k\}} w_{ji} < q_i$. Then $\Omega_p(i)$ is a set of pivotal nodes in group $\Omega(i)$, i.e.

$$\Omega_p(i) = \left\{k \in \Omega(i) \middle| \sum_{j \in \Omega(i) \setminus \{k\}} w_{ji} < q_i\right\}.$$

Generally, every node can have a vector of different attributes depending on the problem statement. These attributes can be estimated by their importance and aggregated to some single value which is its personal threshold $q_i$. For a meaningful comparison of aggregated attributes and weights on nodes these values should be of the same origin. If we do not have individu-al attributes in a network then we can use information from a graph itself. For example, $q_i$ can be a fraction of total in-degree influence on node $i$.

Additionally, critical groups' formation may have some probability, i.e. it is not necessary that some nodes truly want to or can cooperate with each other (depending on the problem statement). This means that some probabilities are attributed to each critical group and they are taken into consideration in the further analysis.

For our Numerical Example 2 the sets of direct neighbors $\overleftarrow{N}_i$ and critical groups when $q_i = 50\%$ of total influence for each node are shown in Table 3. Here we assume that critical groups are formed with probability 1.

**Table 3.** Neighbors and critical groups for Numerical Example 2

| **Node, $i$** | $\overleftarrow{N}_i$ | **Critical groups, $\Omega(i)$, $q$ = 50%** |
|---|---|---|
| 1 | {2, 3, 4, 6} | {2, 3}, {2, 4}, {3, 4}, {4, 6}, {2, 3, 4}, {2, 3, 6}, {3, 4, 6}, {2, 3, 4, 6} |
| 2 | {5, 6, 8} | {5, 6}, {5, 8}, {6, 8}, {5, 6, 8} |
| 3 | {4, 5, 9, 10} | {4, 5}, {4, 10}, {5, 9}, {5, 10}, {9, 10}, {4, 5, 9}, {4, 5, 10}, {5, 9, 10}, {4, 5, 9, 10} |
| 4 | {5, 7, 9, 10} | {5, 7}, {5, 9}, {5, 10}, {7, 9}, {7, 10}, {9, 10}, {5, 7, 9}, {5, 7, 10}, {7, 9, 10}, {5, 7, 9, 10} |
| 5 | ∅ | ∅ |
| 6 | {4, 7} | {7}, {4, 7} |
| 7 | {1, 2, 3} | {2}, {1, 2}, {2, 3}, {1, 2, 3} |
| 8 | {1, 4, 7} | {7}, {1, 7}, {4, 7}, {1, 4 ,7} |
| 9 | {1, 7} | {7}, {1, 7} |
| 10 | {1, 7} | {7}, {1, 7} |

Pivotal members for node 1 when $q_1 = 50\%$ is provided in Table 4.

**Table 4.** Critical groups and pivotal member for node 1

| **Critical groups**, $\Omega(1)$ | **Pivotal members,** $\Omega_p(1)$ |
|---|---|
| {2, 3} | {2, 3} |
| {2, 4} | {2, 4} |
| {3, 4} | {3, 4} |

| **Critical groups**, $\Omega(1)$ | **Pivotal members,** $\Omega_p(1)$ |
|---|---|
| {4, 6} | {4, 6} |
| {2, 3, 4} | ∅ |
| {2, 3, 6} | {2, 3} |
| {3, 4, 6} | {4} |
| {2, 3, 4, 6} | ∅ |

### *3.1. s-long-range interaction index based on paths (s-LRIC index)*

The first approach of the key nodes detection by $s$-LRIC index is based on paths.

Now we construct intensity matrix $C = [c_{ij}]$ with respect to weights $w_{ij}$, thresholds $q_i$ and critical groups $\Omega(j)$ as

$$c_{ij} = \begin{cases} \dfrac{w_{ij}}{\min\limits_{\Omega(j) \subseteq N_j | i \in \Omega_p(j)} \sum_{l \in \Omega(j)} w_{lj}}, if\ i \in \Omega_p(j) \subseteq \overleftarrow{N}_j, \\ 0, i \notin \Omega_p(j) \subseteq \overleftarrow{N}_j, \end{cases} \tag{1}$$

where $\Omega(j) \subseteq \overleftarrow{N}_j$ is a critical group of direct neighbors for node $j$, $\Omega_p(j) \subseteq \Omega(j)$ is a group of pivotal members for $\Omega(j)$.

We consider a critical group with the minimal sum of weights (in denominator) to indicate the maximal possible direct influence of node $i$ on node $j$. Obviously, if $w_{ij} \geq q_j$ then the direct influence of node $i$ on node $j$ is maximal and is equal to 1. Conversely, if node $i$ does not have a direct connection to node $j$ or it does not belong to any critical group then its direct influence is equal to 0. In other cases, if $0 < w_{ij} < q_j$ but node $i$ is pivotal for node $j$ then its direct influence is equal to $c_{ij}$, $0 < c_{ij} < 1$.

Let us construct matrix $C = [c_{ij}]$ for Numerical Example 2 according to formula (1). There are 8 critical groups for node 1 (see Table 4) and, for example, node 3 is pivotal in $\Omega(1) = \{\{2,3\},\{3,4\},\{2,3,6\}\}$; but we consider a critical group with the minimal sum of weights which is $\{2,3\}$; hence, $c_{31} = \frac{w_{31}}{w_{21}+w_{31}} = \frac{300}{500} = 0.6$. Similarly, we define direct influences for other nodes.

Thus, we evaluated the direct influence of the first level on each node in a network. To define the total influence between nodes we need to redesign our graph by the replacement of weights $w_{ij}$ on edges with values of direct influences $c_{ij}$. A new graph of direct influences looks like

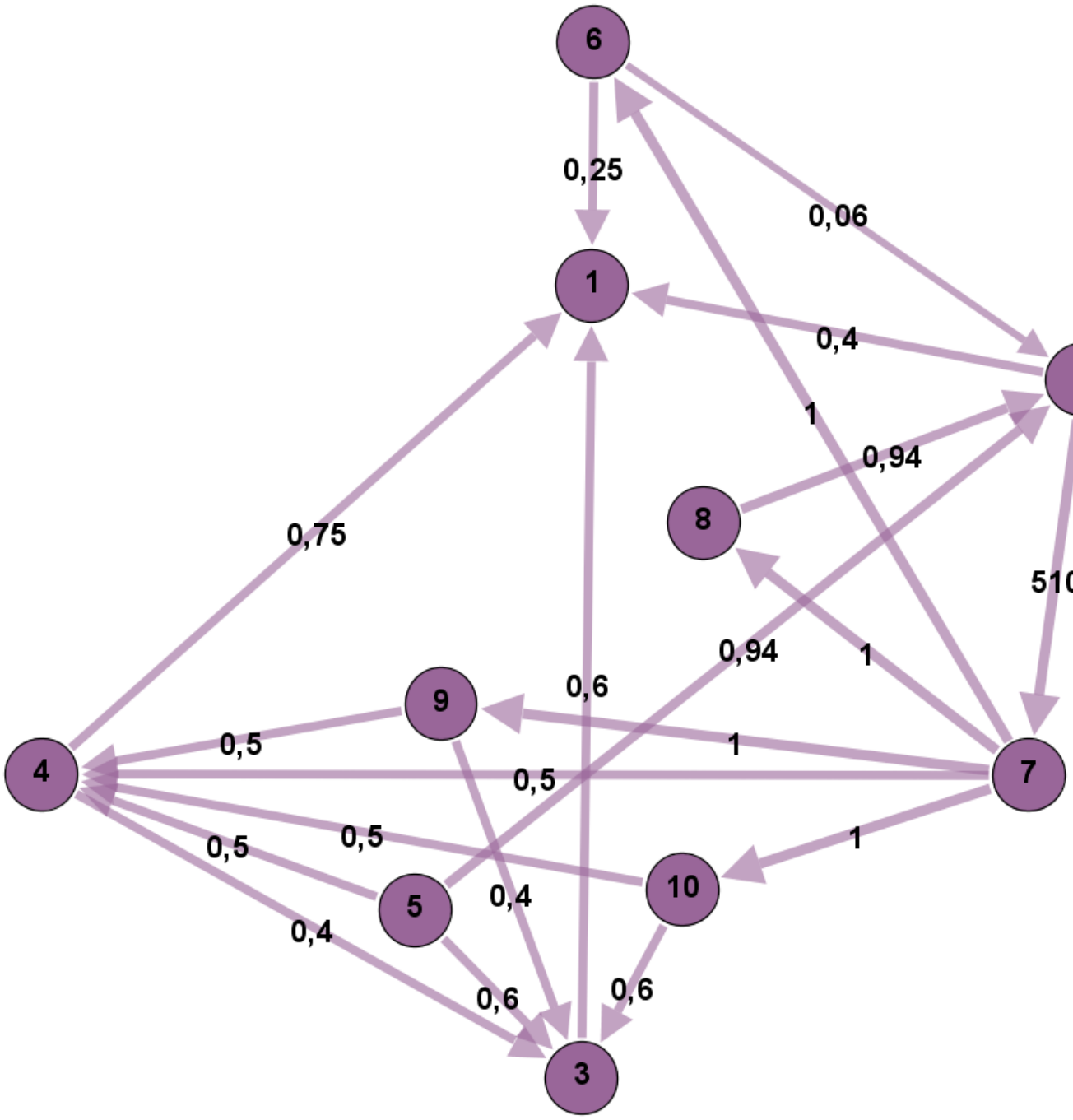

**Fig. 5.** A graph representing matrix $C$ for Numerical Example 2

According to formula (1) node 1 does not influence other nodes with respect to $q_i = 50\%$ of total influence.

Generally, the construction of matrix $C$ is highly related to [Aleskerov et al., 2014] because it requires to consider separately each node $j$ for which we ignore all outgoing edges while other nodes of the graph are assumed as potentially influential on $j$.

To evaluate indirect influences of nodes we need to introduce a concept of a path between a pair of nodes.

*Definition 3*. A path between nodes $i$ and $j$ is a sequence of edges such that the end of one edge is the beginning of the next edge, i.e. $P_k^{ij} = \{(i, l_1), (l_1, l_2), \dots, (l_h, l_{h+1}), \dots, (l_{r-2}, l_{r-1}), (l_{r-1}, j)\}$ is $k$-th path between nodes $i$ and $j$ where $l_h$ is an intermediate node. The number of edges in the sequence is the length of a path (see Fig. 6).

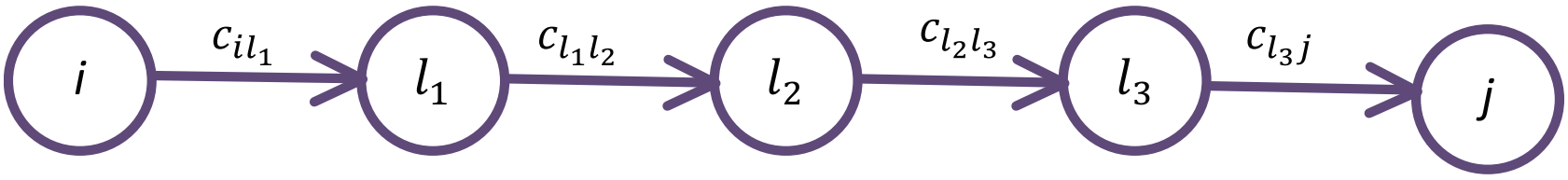

**Fig. 6.** A path in a graph

To analyze the indirect influence of node $i$ on node $j$ we consider all simple paths between them, i.e. paths such that there are no nodes that occur on the path at least twice. For instance, for our Numerical Example 2 there are 4 paths between nodes 7 and 3: {(7, 10), (10, 3)}, {(7, 4), (4, 3)}, {(7, 9), (9, 3)}, {(7, 9), (9, 4), (4, 3)}.

Here we can limit the maximal length of paths with some parameter $s$ because very long paths usually are not representative in terms of indirect influence.

Denote by $P^{ij} = \{P_1^{ij}, \dots, P_m^{ij}\}$ a set of all simple paths between $i$ and $j$, where $m$ is the total number of simple paths, and $n(k) = |P_k^{ij}| \le s$ is equal to the $k$-th path's length. Then the influence of $i$ on $j$ via $k$-th path $P_k^{ij}$ is defined as

$$f(P_k^{ij}) = c_{il_1^k} \times c_{l_1^k l_2^k} \times \dots \times c_{l_{n(k)-1}^k j} \quad (2)$$

or

$$f(P_k^{ij}) = \min\left(c_{il_1^k}, c_{l_1^k l_2^k}, \dots, c_{l_{n(k)-1}^k j}\right) \quad (3)$$

where $i, l_1^k, l_2^k, \dots, l_{n(k)-1}^k, j$ is an ordered sequence of nodes in the $k$-th path.

According to the formula (2) the influence of node $i$ on node $j$ through the $k$-th path $P_k^{ij}$ is calculated as the aggregate value of direct influences between nodes which lie in this path. The formula (3) can be interpreted as the $k$-th path capacity of the influence (we cannot influence through the $k$-th path more than the minimal value of the influence is allowed on this path).

After we considered the influence of node *i* on node *j* through all paths of length less than or equal to *s* (formula (2) or (3)) we need to aggregate the total influence of node *i* on node *j*. We propose three ways of the aggregation of the possible influence; the aggregated results form new matrix $C^* = [c_{ij}^*]$:

1. *The total influence via the sum of possible influences*

$$c_{ij}^*(s) = \min\left\{ \sum_{k:\left|P_k^{ij}\right| \le s} f\left(P_k^{ij}\right), \quad 1 \right\} \tag{4}$$

2. *The total influence via maximum possible influence*

$$c_{ij}^*(s) = \max_{k:\left|P_k^{ij}\right| \le s} f(P_k^{ij}) \tag{5}$$

3. *The total influence via the threshold aggregation*

The model of the threshold aggregation was proposed in [Aleskerov et al., 2007]. Each node in a graph with *n* nodes is evaluated by *n* grades that may have $m$ different values. Then for each node $k$ we calculate values $v_1(k), \dots v_m(k)$ where $v_i(k)$ is the number of $i$-th grades that node $k$ received, *i=1,…,m*. According to the threshold rule node *x* *V*-dominates node *y* if $v_1(x) < v_1(y)$ or $\exists d \le m: \forall h < d\ v_h(x) = v_h(y)$ and $v_d(x) < v_d(y)$. In other words, firstly we compare the number of the worst grades; if they are equal then we compare the number of the second worst grades, etc. If some node is not *V*-dominated by other nodes then this node is considered as the best one.

Considering the threshold rule as one of possible ways on how the indirect influence can be evaluated, we propose the following aggregation procedure

$$c_{ij}^*(s) = f\left(P_z^{ij}\right) \tag{6}$$

where

$$z = \operatorname*{argmin}_{k:\, n(k) \le s} v(P_k^{ij}) \tag{7}$$

and

$$v\left(P_k^{ij}\right) = \sum_{l=1}^{m} v_l\left(P_k^{ij}\right) * (s+1)^{m-l} + s - n(k) \tag{8}$$

Formula (6) – (8) are identical to the threshold rule [Aleskerov et al. 2010].

Hence, we construct a matrix $C^*(s) = [c_{ij}^*(s)]$ where $c_{ij}^*(s)$ is a total influence of node $i$ on node $j$ with respect to paths of length less than $s$. Note that if there are no paths between nodes $i$ and $j$ then $\forall s\ c_{ij}^*(s) = 0$.

As a result, we propose two methods of path power estimation and three methods of the aggregation of possible influence. Hence, we can assume that there are 6 ways of the estimation of long-range interactions in a graph. However, not all combinations of (2) – (3) and (4) – (6) are reasonable. Thus, we propose the following combinations (see Table 5):

- (2) – (4): an influence of $i$ on $j$ goes through all paths with respect to all layers in these paths;
- (2) – (5): an influence of $i$ on $j$ goes through a maximal path with respect to all layers in this path;
- (2) – (6): an influence of $i$ on $j$ goes through the best paths according to the threshold rule with respect to all layers in these paths;
- (3) – (5): an influence of $i$ on $j$ goes through a maximal path with respect to one minimal layer in this path;
- (3) – (6): an influence of $i$ on $j$ goes through the best paths according to the threshold rule with respect to one minimal layer in this path.

The combination (3) – (4) is not very reasonable because, firstly, for every path we evaluate the capacity of the influences which can confine on one edge for different paths; then we sum up these influences, which means that we may consider the same influence several times.

**Table 5.** Possible combinations of methods for indirect influence

<table>
<tr><th colspan="2"></th><th colspan="3">Paths aggregation</th></tr>
<tr><th colspan="2"></th><th>Sum of paths influences</th><th>Maximal path influence</th><th>Threshold rule</th></tr>
<tr><th rowspan="2">Path influence</th><th>Multiplication of direct influence</th><td>SumPaths</td><td>MaxPath</td><td>MultT</td></tr>
<tr><th>Minimal direct influence</th><td>–</td><td>MaxMin</td><td>MaxT</td></tr>
</table>

For our Numerical Example 2 we evaluate the indirect influence of node 7 on node 1 through all existing paths in the graph (see Table 6).

**Table 6.** The indirect influence of node 7 on node 1 via paths

| ID | Simple Paths | Multiplication of paths' influences (2) | Minimal direct influence (3) |
|---|---|---|---|
| 1 | $7 \xrightarrow{0,5} 4 \xrightarrow{0,75} 1$ | 0,375 | 0,5 |
| 2 | $7 \xrightarrow{0,5} 4 \xrightarrow{0,4} 3 \xrightarrow{0,6} 1$ | 0,12 | 0,4 |
| 3 | $7 \xrightarrow{1} 6 \xrightarrow{0,25} 1$ | 0,25 | 0,25 |
| 4 | $7 \xrightarrow{1} 6 \xrightarrow{0,06} 2 \xrightarrow{0,4} 1$ | 0,024 | 0,06 |
| 5 | $7 \xrightarrow{1} 8 \xrightarrow{0,94} 2 \xrightarrow{0,4} 1$ | 0,376 | 0,4 |
| 6 | $7 \xrightarrow{1} 9 \xrightarrow{0,4} 3 \xrightarrow{0,6} 1$ | 0,24 | 0,4 |
| 7 | $7 \xrightarrow{1} 9 \xrightarrow{0,5} 4 \xrightarrow{0,75} 1$ | 0,375 | 0,5 |
| 8 | $7 \xrightarrow{1} 10 \xrightarrow{0,6} 3 \xrightarrow{0,6} 1$ | 0,36 | 0,6 |
| 9 | $7 \xrightarrow{1} 10 \xrightarrow{0,5} 4 \xrightarrow{0,75} 1$ | 0,375 | 0,5 |

If we do not limit the length of a path with parameter *s* then there are 9 ways of the influence of node 7 on node 1; otherwise, if we limit the length of a path with, for example, $s = 2$, then we will consider the influence of 7 on 1 only through 2 paths ($1^{st}$ and $3^{rd}$ ones).

To compare different paths by the threshold rule the following grades of direct influence are proposed.

*Grades:*

**0.** $0 \leq c_{ij} < 0.2;$

**1.** $0.2 \leq c_{ij} < 0.5;$

**2.** $0.5 \leq c_{ij} < 0.7;$

**3.** $0.7 \leq c_{ij} \leq 1.$

Now we can define the path between node 7 and node 1 according to the threshold rule. Note that for the threshold rule the values on the edges are equal to the grades which were proposed above. The results are provided in Table 7.

**Table 7.** Paths aggregation by the threshold rule, *s*=3, *m*=4

| **ID, *k*** | **Path** | **Path (grades on edges)** | **Paths influence, $v(P_k^{71})$** |
|---|---|---|---|
| 1 | $7 \xrightarrow{0,5} 4 \xrightarrow{0,75} 1$ | $7 \xrightarrow{2} 4 \xrightarrow{3} 1$ | 21* |
| 2 | $7 \xrightarrow{0,5} 4 \xrightarrow{0,4} 3 \xrightarrow{0,6} 1$ | $7 \xrightarrow{2} 4 \xrightarrow{1} 3 \xrightarrow{2} 1$ | 96 |
| 3 | $7 \xrightarrow{1} 6 \xrightarrow{0,25} 1$ | $7 \xrightarrow{3} 6 \xrightarrow{1} 1$ | 69 |
| 4 | $7 \xrightarrow{1} 6 \xrightarrow{0,06} 2 \xrightarrow{0,4} 1$ | $7 \xrightarrow{3} 6 \xrightarrow{0} 2 \xrightarrow{1} 1$ | 324 |
| 5 | $7 \xrightarrow{1} 8 \xrightarrow{0,94} 2 \xrightarrow{0,4} 1$ | $7 \xrightarrow{3} 8 \xrightarrow{3} 2 \xrightarrow{1} 1$ | 72 |
| 6 | $7 \xrightarrow{1} 9 \xrightarrow{0,4} 3 \xrightarrow{0,6} 1$ | $7 \xrightarrow{3} 9 \xrightarrow{1} 3 \xrightarrow{2} 1$ | 84 |
| 7 | $7 \xrightarrow{1} 9 \xrightarrow{0,5} 4 \xrightarrow{0,75} 1$ | $7 \xrightarrow{3} 9 \xrightarrow{2} 4 \xrightarrow{3} 1$ | 24 |
| 8 | $7 \xrightarrow{1} 10 \xrightarrow{0,6} 3 \xrightarrow{0,6} 1$ | $7 \xrightarrow{3} 10 \xrightarrow{2} 3 \xrightarrow{2} 1$ | 36 |
| 9 | $7 \xrightarrow{1} 10 \xrightarrow{0,5} 4 \xrightarrow{0,75} 1$ | $7 \xrightarrow{3} 10 \xrightarrow{2} 4 \xrightarrow{3} 1$ | 24 |

* is the path chosen by the threshold rule.

Thus, there are 9 possible ways how node 7 influences node 1. Let us now aggregate this information into a single value by different methods. The overall results are provided in Table 8.

**Table 8.** The total influence of node to node 1 by different methods

| **Method** | **Considered paths IDs** | **Influence** |
|---|---|---|
| SumPaths | 1–9 | 1 |
| MaxPath | 5 | 0,376 |
| MaxMin | 8 | 0,6 |
| MultT | 1 | 0,375 |
| MaxT | 1 | 0,5 |

Similarly, we can estimate the influence of any other elements and construct the matrix $C^*$ according to different methods. The results are provided in Tables 9–13.

**Table 9.** Matrix $C^*$ for Numerical Example 2, SumPaths

| | 1 | 2 | 3 | 4 | 5 | 6 | 7 | 8 | 9 | 10 |
|---|---|---|---|---|---|---|---|---|---|---|
| **1** | 0 | 0 | 0 | 0 | 0 | 0 | 0 | 0 | 0 | 0 |
| **2** | 1 | 0 | 1 | 1 | 0 | 1 | 1 | 1 | 1 | 1 |
| **3** | 0,6 | 0 | 0 | 0 | 0 | 0 | 0 | 0 | 0 | 0 |
| **4** | 0,99 | 0 | 0,4 | 0 | 0 | 0 | 0 | 0 | 0 | 0 |
| **5** | 1 | 0,942 | 1 | 1 | 0 | 0,942 | 0,942 | 0,942 | 0,942 | 0,942 |
| **6** | 0,381 | 0,058 | 0,093 | 0,087 | 0 | 0 | 0,058 | 0,058 | 0,058 | 0,058 |
| **7** | 1 | 1 | 1 | 1 | 0 | 1 | 0 | 1 | 1 | 1 |
| **8** | 1 | 0,942 | 1 | 1 | 0 | 0,942 | 0,942 | 0 | 0,942 | 0,942 |
| **9** | 0,735 | 0 | 0,6 | 0,5 | 0 | 0 | 0 | 0 | 0 | 0 |
| **10** | 0,855 | 0 | 0,8 | 0,5 | 0 | 0 | 0 | 0 | 0 | 0 |

**Table 10.** Matrix $C^*$ for Numerical Example 2, MaxPath

| | 1 | 2 | 3 | 4 | 5 | 6 | 7 | 8 | 9 | 10 |
|---|---|---|---|---|---|---|---|---|---|---|
| **1** | 0 | 0 | 0 | 0 | 0 | 0 | 0 | 0 | 0 | 0 |
| **2** | 0,4 | 0 | 0,6 | 0,5 | 0 | 1 | 1 | 1 | 1 | 1 |
| **3** | 0,6 | 0 | 0 | 0 | 0 | 0 | 0 | 0 | 0 | 0 |
| **4** | 0,75 | 0 | 0,4 | 0 | 0 | 0 | 0 | 0 | 0 | 0 |
| **5** | 0,377 | 0,942 | 0,6 | 0,5 | 0 | 0,942 | 0,942 | 0,942 | 0,942 | 0,942 |
| **6** | 0,25 | 0,058 | 0,035 | 0,029 | 0 | 0 | 0,058 | 0,058 | 0,058 | 0,058 |
| **7** | 0,377 | 0,942 | 0,6 | 0,5 | 0 | 1 | 0 | 1 | 1 | 1 |
| **8** | 0,377 | 0,942 | 0,565 | 0,471 | 0 | 0,942 | 0,942 | 0 | 0,942 | 0,942 |
| **9** | 0,375 | 0 | 0,4 | 0,5 | 0 | 0 | 0 | 0 | 0 | 0 |
| **10** | 0,375 | 0 | 0,6 | 0,5 | 0 | 0 | 0 | 0 | 0 | 0 |

**Table 11.** Matrix *C** for Numerical Example 2, MaxMin

| | 1 | 2 | 3 | 4 | 5 | 6 | 7 | 8 | 9 | 10 |
|---|---|---|---|---|---|---|---|---|---|---|
| **1** | 0 | 0 | 0 | 0 | 0 | 0 | 0 | 0 | 0 | 0 |
| **2** | 0,6 | 0 | 0,6 | 0,5 | 0 | 1 | 1 | 1 | 1 | 1 |
| **3** | 0,6 | 0 | 0 | 0 | 0 | 0 | 0 | 0 | 0 | 0 |
| **4** | 0,75 | 0 | 0,4 | 0 | 0 | 0 | 0 | 0 | 0 | 0 |
| **5** | 0,6 | 0,942 | 0,6 | 0,5 | 0 | 0,942 | 0,942 | 0,942 | 0,942 | 0,942 |
| **6** | 0,25 | 0,058 | 0,058 | 0,058 | 0 | 0 | 0,058 | 0,058 | 0,058 | 0,058 |
| **7** | 0,6 | 0,942 | 0,6 | 0,5 | 0 | 1 | 0 | 1 | 1 | 1 |
| **8** | 0,6 | 0,942 | 0,6 | 0,5 | 0 | 0,942 | 0,942 | 0 | 0,942 | 0,942 |
| **9** | 0,5 | 0 | 0,4 | 0,5 | 0 | 0 | 0 | 0 | 0 | 0 |
| **10** | 0,6 | 0 | 0,6 | 0,5 | 0 | 0 | 0 | 0 | 0 | 0 |

**Table 12.** Matrix *C** for Numerical Example 2, MultT

| | 1 | 2 | 3 | 4 | 5 | 6 | 7 | 8 | 9 | 10 |
|---|---|---|---|---|---|---|---|---|---|---|
| **1** | 0 | 0 | 0 | 0 | 0 | 0 | 0 | 0 | 0 | 0 |
| **2** | 0,375 | 0 | 0,6 | 0,5 | 0 | 1 | 1 | 1 | 1 | 1 |
| **3** | 0,6 | 0 | 0 | 0 | 0 | 0 | 0 | 0 | 0 | 0 |
| **4** | 0,75 | 0 | 0,4 | 0 | 0 | 0 | 0 | 0 | 0 | 0 |
| **5** | 0,375 | 0,942 | 0,6 | 0,5 | 0 | 0,942 | 0,942 | 0,942 | 0,942 | 0,942 |
| **6** | 0,25 | 0,058 | 0,035 | 0,029 | 0 | 0 | 0,058 | 0,058 | 0,058 | 0,058 |
| **7** | 0,375 | 0,942 | 0,6 | 0,5 | 0 | 1 | 0 | 1 | 1 | 1 |
| **8** | 0,353 | 0,942 | 0,565 | 0,471 | 0 | 0,942 | 0,942 | 0 | 0,942 | 0,942 |
| **9** | 0,375 | 0 | 0,4 | 0,5 | 0 | 0 | 0 | 0 | 0 | 0 |
| **10** | 0,375 | 0 | 0,6 | 0,5 | 0 | 0 | 0 | 0 | 0 | 0 |

**Table 13.** Matrix $C*$ for Numerical Example 2, MaxT

| | 1 | 2 | 3 | 4 | 5 | 6 | 7 | 8 | 9 | 10 |
|---|---|---|---|---|---|---|---|---|---|---|
| **1** | 0 | 0 | 0 | 0 | 0 | 0 | 0 | 0 | 0 | 0 |
| **2** | 0,5 | 0 | 0,6 | 0,5 | 0 | 1 | 1 | 1 | 1 | 1 |
| **3** | 0,6 | 0 | 0 | 0 | 0 | 0 | 0 | 0 | 0 | 0 |
| **4** | 0,75 | 0 | 0,4 | 0 | 0 | 0 | 0 | 0 | 0 | 0 |
| **5** | 0,5 | 0,942 | 0,6 | 0,5 | 0 | 0,942 | 0,942 | 0,942 | 0,942 | 0,942 |
| **6** | 0,25 | 0,058 | 0,058 | 0,058 | 0 | 0 | 0,058 | 0,058 | 0,058 | 0,058 |
| **7** | 0,5 | 0,942 | 0,6 | 0,5 | 0 | 1 | 0 | 1 | 1 | 1 |
| **8** | 0,5 | 0,942 | 0,6 | 0,5 | 0 | 0,942 | 0,942 | 0 | 0,942 | 0,942 |
| **9** | 0,5 | 0 | 0,4 | 0,5 | 0 | 0 | 0 | 0 | 0 | 0 |
| **10** | 0,5 | 0 | 0,6 | 0,5 | 0 | 0 | 0 | 0 | 0 | 0 |

After we constructed the matrix of node-to-node influence $C^*(s)$ we can estimate the influence of a node within the whole graph. The aggregation of a matrix to a single vector of the influence $\tilde{c}(s)$ depends on the problem statement. Generally, we can use some pre-defined attributes of nodes or any other factors. For a network of influence we can, for example, estimate a relative independence of a node on other nodes, i.e. a weight of a node is higher if a smaller number of nodes influences this node with respect to the total influence through a graph.

For Numerical Example 2 we can use the following aggregation approach: a weight of node $i$ is its relative influence on other nodes with respect to the whole graph influence, i.e.

$$u^i = \frac{\sum_j w_{ij}}{\sum_k \sum_j w_{kj}}.$$

The higher the influence of a node on other nodes the higher its weight is. Then the vector of influence $\tilde{c}(s)$ is

$$\tilde{c}(s) = C^* \cdot u,$$

where $u = (u^1, \dots, u^{10})$.

For Numeric Example 2 vector $u$ is calculated in Table 14.

**Table 14.** Individual weights of nodes

| **Node, $i$** | 1 | 2 | 3 | 4 | 5 | 6 | 7 | 8 | 9 | 10 |
|---|---|---|---|---|---|---|---|---|---|---|
| **Weight, $u^i$** | 0,17 | 0,079 | 0,054 | 0,145 | 0,115 | 0,017 | 0,254 | 0,054 | 0,05 | 0,061 |

In Table 15 we compare results that are obtained by proposed approaches (5 versions).

**Table 15.** Aggregated values for Numerical Example 2

| **Node** | **LRIC** | | | | |
|---|---|---|---|---|---|
| | **SumPaths** | **MaxPath** | **MaxMin** | **MultT** | **MaxT** |
| 1 | 0 | 0 | 0 | 0 | 0 |
| 2 | **0,806** | **0,610** | **0,644** | **0,606** | **0,627** |
| 3 | 0,102 | 0,102 | 0,102 | 0,102 | 0,102 |
| 4 | 0,190 | 0,149 | 0,149 | 0,149 | 0,149 |
| 5 | **0,855** | **0,655** | **0,693** | **0,654** | **0,676** |
| 6 | 0,112 | 0,078 | 0,083 | 0,078 | 0,083 |
| 7 | 0,631 | 0,426 | 0,464 | 0,425 | 0,447 |
| 8 | **0,804** | **0,598** | **0,642** | **0,594** | **0,625** |
| 9 | 0,230 | 0,158 | 0,179 | 0,158 | 0,179 |
| 10 | 0,261 | 0,169 | 0,207 | 0,169 | 0,190 |

We can see that all versions of the LRIC index detect nodes 2, 5, 8 as the most influential.

### *3.2. s-long-range interactions index based on simulations*

The second approach of key node detection by *s*-LRIC index is based on simulations. The idea of this method is as follows: suppose that we unite some group of nodes and influence other nodes by this group; some influenced nodes join the first group with respect to their thresholds; the expanded group again influences the rest of nodes and some of them join this expanded group. We continue this procedure step by step until all nodes join the first group or there are no more nodes that are affected by final expanded group. Additionally, we can limit the number of steps with some parameter *s* and we can stop the procedure when we reach this limit. On the next stage, we unite another group of nodes and follow their effect of the influence on other nodes. When we study all groups that can be chosen on the first step, for each node we summarize the information about nodes that joined the first group afterwards due to the fact that this node was in the first group. Thus, we get node-to-node influences and this information can be aggregated into a single vector.

Formally, we have the network-graph $G(V,E)$, where $V=\{1,\dots,n\}$ is a set of nodes, $|V|=N$, $E=\{(i,j),\ \ i,j\in V\}$ is a set of weighted edges, and $w_{ij}$ is a weight of edge $(i,j)$. Let us illustrate the approach using Numerical Example 2.

Firstly, we construct matrix $C=[c_{ij}]$ with respect to weights $w_{ij}$ and predefined thresholds $q_i$ (a level of absolute influence)

$$c_{ij}=\min\left\{\frac{w_{ij}}{q_i},1\right\} \tag{9}$$

Matrix $C$ indicates the share of the influence of node $j$ on node $i$ with respect to threshold $q_i$. For Numerical Example 2 consider thresholds $q_i=50\%$ of the total influence on node *i*. A graph of relative influences for Numerical Example 2 is provided on Fig. 7.

Here we introduce a concept of minimal direct critical groups of node *i*.

*Definition 4.* A group of neighbors of node $i$ $\Omega^d(i)\subseteq\overleftarrow{N}_i$ is a minimal direct critical group if $\sum_{j\in\Omega^d(i)}c_{ji}\geq 1$and $\forall k\in\Omega^d(i)\sum_{j\in\Omega^d(i)\setminus\{k\}}c_{ji}<1$, i.e. according to Definition 2 all members are pivotal.

In Table 16 nodes' neighbors and minimal direct critical groups are represented for Numerical Example 2.

Now we activate some group of nodes to follow its effect on the graph. It is unreasonable to choose a group that does not contain any minimal direct critical group, because such groups are not influential. For instance, for Numerical Example 2 it is useless to choose nodes 1, 3 and 10 because they do not influence other nodes enough (with respect to thresholds). Contrary, if we choose nodes 2 and 7, then they influence nodes 6, 8, 9, 10 (because of node 7); expanded group {2, 6, 7, 8, 9, 10} influences node 4 (because of critical groups {7, 9}, {7, 10} or {9, 10}). A new expanded group {2, 4, 6, 7, 8, 9, 10} influences node 1 (because of the critical groups {2, 4} or {4, 6}) and node 3 (because of the critical group {4, 10}). Final expanded group {1, 2, 3, 4, 6, 7, 8, 9, 10} does not influences the rest node 5 and we stop at this stage. Hence, nodes 2 and 7 explicitly affect all nodes except node 5.

However, if we limit the number of steps with parameter *s* and assume *s* = 2 then nodes 2 and 7 will not influence nodes 1 and 3 either.

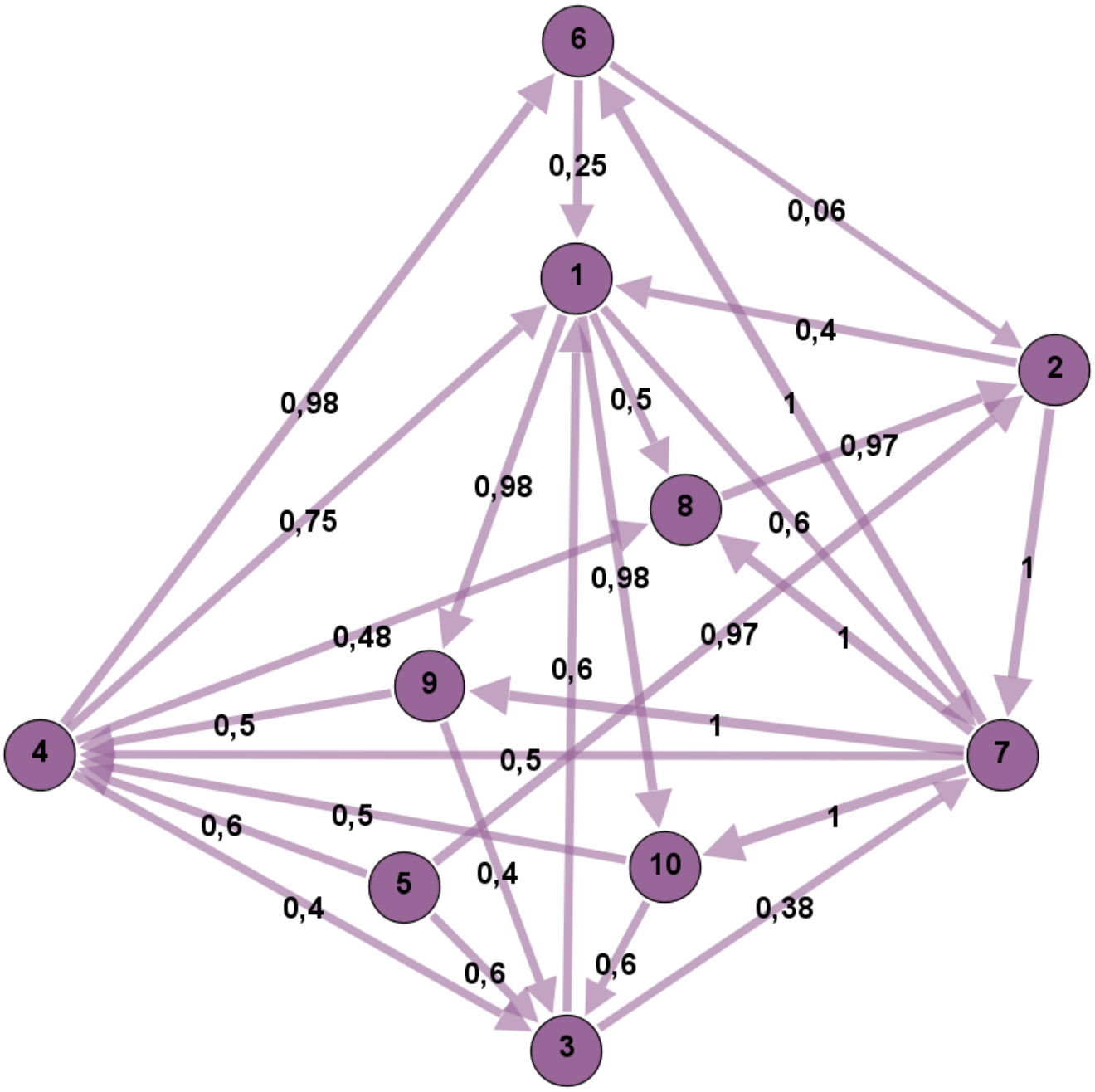

**Fig. 7.** Influence intensities for Numerical Example 2

**Table 16.** Minimal direct critical groups for Numerical Example 2

| **Node, *i*** | **Neighbors, $\overleftarrow{N}_i$** | **Minimal direct critical groups, $\Omega^d(i)$** |
|---|---|---|
| 1 | {2, 3, 4, 6} | {2, 3}, {2, 4}, {3, 4}, {4, 6} |
| 2 | {5, 6, 8} | {5, 6}, {5, 8}, {6, 8} |
| 3 | {4, 5, 9, 10} | {4, 5}, {4, 10}, {5, 9}, {5, 10} |
| 4 | {5, 7, 9, 10} | {5, 7}, {5, 9}, {5, 10}, {7, 9}, {7, 10}, {9, 10} |
| 5 | ∅ | ∅ |
| 6 | {4, 7} | {7} |
| 7 | {1, 2, 3} | {2} |
| 8 | {1, 4, 7} | {7} |
| 9 | {1, 7} | {7} |
| 10 | {1, 7} | {7} |

Let $K_l^0$ be some group of nodes that are chosen on the first step, and $K_l^F$ be a final expanded group that is obtained from $K_l^0$. When we consider all possible $K_l^0$ (in the worst case we need to consider $2^N - 2$ groups of nodes) we can form resulting matrix of node-to-node influence $C^*(s) = [c_{ij}^*(s)]$ where

$$c_{ij}^*(s) = \frac{\left|\{l: j \in K_l^F \mid i \in K_l^0,\ i \in \Omega^d(j)\}\right|}{\left|\{l: i \in K_l^0\}\right| - \left|\{l: i,j \in K_l^0\}\right|}. \tag{10}$$

In other words, the influence of node $i$ on node $j$ is the number of times node $j$ hits into final broaden group when node $i$ is chosen on the first step and node $i$ is contained in minimal direct critical group of node $j$ divided by the number of times node $i$ is chosen on the first step but not together with node $j$. Apparently, if $i \notin \Omega^d(j)$ then $c_{ij}^*(s) = 0$.

Note that if we consider all possible groups on the first step then each node is met there in a half of the instances, and any pair of nodes is met in a quarter of the instances. Hence, denominator in formula (10) is equal to $\frac{2^N-2}{2} - \frac{2^N-2}{4} = 2^{N-1} - 2^{N-2} - 2^{-1}$.

The interpretation of the matrix $C^*(s)$ is straightforward. If value $c_{ij}^*$ is close to 1 then node $i$ tremendously influences node $j$. On the contrary, if value $c_{ij}^*$ is close to 0 then node $i$ poorly influences node $j$. Again, we can aggregate node-to-node influences to a single vector as was described above.

The results of simulation approach for Numerical Example 2 are presented in Table 17.

**Table 17.** Simulations results for Numerical Example 2

| **Node, *i*** | 1 | 2 | 3 | 4 | 5 | 6 | 7 | 8 | 9 | 10 |
|---|---|---|---|---|---|---|---|---|---|---|
| **LRIC (Simul)** | 0 | **0,239** | 0,01 | 0,047 | **0,16** | 0,119 | **0,187** | 0,124 | 0,053 | 0,062 |

Simulations detect nodes 2, 5 and 7 as the most influential in the graph. We can see that this approach detects pivotal node 7 that is chosen by most of the classical measures and two implicit influential nodes 2 and 5 that are chosen by all versions of LRIC index.

One of the key advantages of this approach is that it accurately takes into account all chain reactions of a graph, so-called domino or contagion effect.

On the other hand, as this method is highly memory consuming (in the worst case we need to consider all combinations) this brings up the question of the limitation of considered combinations, i.e. which node's combinations should be chosen on the first stage and which ones can be ignored. Usually, in most real-life situations the formation of nodes groups has some probability; this means that depending on the problem statement it is not always practical to consider combinations with a low probability to emerge.

If there are no natural limitations on the number of combinations then it is reasonable to introduce some general limitations, for example, we can limit the size of $K_l^0$ with some predefined parameter $n_0$, i.e. $|K_l^0| \leq n_0$; or we can set limits on the number of combinations and chain reactions with parameter *s,* etc.

Let us compare our results with classical centrality measures and SRIC index. In Table 18 classical centrality measures and SRIC index are calculated for Numerical Example 2.

**Table 18.** Classical centrality measures and SRIC index for Numerical Example 2

| Centrality / Node | $C^{w\ in-deg}$ | $C^{w\ out-deg}$ | $C^{w\ deg}$ | $C^{w\ deg-diff}$ | $C^{btw}$ | $C^{cl}$ (*) | $C^{PageRank}$ | $C^{eig}$ | $C^{Bonacich}$ | SRIC |
|---|---|---|---|---|---|---|---|---|---|---|
| 1 | 1000 | **1530** | **2530** | **530** | **11** | 0,00069 | **0,176** | **0,826** | –1,034 | 0 |
| 2 | 1000 | 710 | 1710 | –290 | **18** | 0,00124 | 0,105 | 0,565 | –1,034 | **0,315** |
| 3 | 1000 | 490 | 1490 | –510 | 1 | 0,00040 | 0,109 | 0,494 | –1,034 | 0,041 |
| 4 | 1000 | **1305** | **2305** | 305 | 11 | 0,00060 | **0,109** | **0,717** | –1,034 | **0,123** |
| 5 | 0 | 1035 | 1035 | **1035** | 0 | 0,00103 | 0,015 | 0,303 | –1,330 | 0,102 |
| 6 | 1000 | 155 | 1155 | –845 | 0 | **0,00025** | 0,077 | 0,494 | –0,887 | 0,009 |
| 7 | 1000 | **2290** | **3290** | **1290** | **22** | 0,00117 | **0,144** | **1** | –1,182 | **0,264** |
| 8 | 1000 | 485 | 1485 | –515 | 9 | 0,00099 | 0,084 | 0,586 | **–0,443** | 0,043 |
| 9 | 1000 | 450 | 1450 | –550 | 1,5 | **0,00030** | 0,090 | 0,601 | **–0,887** | 0,042 |
| 10 | 1000 | 550 | 1550 | –450 | 6,5 | **0,00031** | 0,090 | 0,626 | **–0,887** | 0,059 |

(*) Closeness centrality: inverse average maximal outflow ⇒ low values are more significant.

We can see that classical centrality measures detect nodes 1, 4, 7 as the most influential while SRIC index detects nodes 2, 4, 7. None of these indices (except Bonacich) indicate node 8 as the influential one, however it affects node 2 very strong which in its turn influences nodes 1 and 7. Moreover, node 5, which is not considered by classical measures and SRIC index, influences nodes 2, 3 and 4, which are very influential too. Additionally, node 1 does not play a significant role in the graph because its outgoing intensities are relatively small. Hence, classical centrality measures are not very appropriate for the influence estimation; SRIC index can be applied to small graphs with short paths between nodes (due to the fact that SRIC index consider only one layer between nodes). LRIC index can identify hidden nodes that are very important in terms of the influence.

In order to compare rankings, we used a correlation analysis. Since the position in the ranking is a rank variable, to assess the consistency of different orderings other than traditional Pearson coefficient rank correlation coefficients should be used. In our work it is applied the idea of Kendall metrics [Kendall, 1970], that counts the number of pairwise disagreements between two ranking lists. We have used Goodman and Kruskal γ rank coefficient as well, which shows the similarity of the orderings of the data when ranked by each of the quantities [Goodman, Kruskal, 1954]. This coefficient looks as $\gamma = \frac{N_S - N_D}{N_S + N_D}$, where $N_S$ is the number of pairs of cases ranked in the same order on both variables (number of concordant pairs) and $N_D$ is the number of pairs of cases ranked in reversed order on both variables (number of reversed pairs).

The results are provided below (see Tables 22–23).

## 4. Computational complexity of centrality measures

We will discuss here a computational complexity of classical and proposed centrality measures. An important issue of classical centrality measures is the scope of information they aggregate: the more information about nodes they consider, the more difficult it is to calculate them for large networks.

Since in-degree, out-degree and degree centralities calculate the number of edges for each node, the computational complexity of these measures is linear. In order to calculate the closeness and betweenness centralitities, the shortest paths between all node-pairs should be considered (total number is $|V| \cdot (|V| - 1)$). The fastest known single-source shortest-path algorithm is the Dijkstra's algorithm proposed in [Dijkstra, 1959] that has a worst case performance equal to $O(|E| + |V| \cdot \log |V|)$. Thus, closeness and betweenness centralities are more difficult to calculate for large networks since they require a polynomial time. As for the eigenvector centrality and its counterparts, these measures have a polynomial computational complexity since they require to compute the eigenvector of adjacency matrix $A$. However, many approximate algorithms with a low computational complexity that implement the main idea of these centrality measures were proposed and integrated in standard software packages.

As for the Myerson value, this measure requires to consider all subgraphs (the total number is $2^{|V|} - 1$) which is impossible to do for a large number of nodes. There have been developed some approximate algorithms to calculate the Shapley value (which provides a basis for the Myerson value); in [Fatima et al., 2008] there is an extended benchmark study of some approximation methods.

As to proposed models, SRIC and LRIC consider all possible pivotal groups for each node the total number of which in the worst case is $2^{|V|-1} - 1$. Distant interactions require the enumeration of all simple paths in a graph, which can reach more than $|V|!$ operating cycles. However, some simplifications can be introduced, as the limitation of the size of groups and path length. Simulations require the consideration of all subgroups of nodes; however, it is not rational to enumerate all of them, because not all subgroups are influential.

The calculation of classical centrality measures was performed in R 3.2.2 software package with the use of embedded functions ("igraph" package). The calculation of proposed measures was produced with the help of MATLAB R2015.

**Table 22.** Kendall τ-coefficient

| | $C^{w\ in-deg}$ | $C^{w\ out-deg}$ | $C^{w\ deg}$ | $C^{w\ deg-diff}$ | $C^{btw}$ | $C^{cl}$ | $C^{PageRank}$ | $C^{eig}$ | $C^{Bonacich}$ | **SRIC** | **SumPaths** | **MaxPath** | **MaxMin** | **MultT** | **MaxT** | **Simul** |
|---|---|---|---|---|---|---|---|---|---|---|---|---|---|---|---|---|
| $C^{w\ in-deg}$ | – | –0.15 | 0.45 | –0.35 | 0.41 | 0.25 | 0.46 | 0.45 | 0.50 | –0.15 | –0.45 | –0.45 | –0.45 | –0.45 | –0.45 | –0.25 |
| $C^{w\ out-deg}$ | | – | 0.73 | 0.91 | 0.52 | –0.51 | 0.52 | 0.45 | –0.60 | 0.42 | 0.11 | 0.16 | 0.16 | 0.16 | 0.16 | 0.02 |
| $C^{w\ deg}$ | | | – | 0.64 | 0.75 | –0.33 | 0.80 | 0.72 | –0.30 | 0.24 | –0.16 | –0.11 | –0.11 | –0.11 | –0.11 | –0.16 |
| $C^{w\ deg-diff}$ | | | | – | 0.43 | –0.60 | 0.43 | 0.36 | –0.70 | 0.42 | 0.20 | 0.24 | 0.24 | 0.24 | 0.24 | 0.11 |
| $C^{btw}$ | | | | | – | –0.48 | 0.54 | 0.64 | –0.18 | 0.43 | 0.11 | 0.16 | 0.16 | 0.16 | 0.16 | 0.11 |
| $C^{cl}$ | | | | | | – | –0.21 | –0.14 | 0.45 | –0.56 | –0.42 | –0.47 | –0.47 | –0.47 | –0.47 | –0.42 |
| $C^{PageRank}$ | | | | | | | – | 0.60 | –0.36 | 0.02 | –0.39 | –0.34 | –0.34 | –0.34 | –0.34 | –0.39 |
| $C^{eig}$ | | | | | | | | – | –0.08 | 0.14 | –0.23 | –0.23 | –0.23 | –0.23 | –0.23 | –0.23 |
| $C^{Bonacich}$ | | | | | | | | | – | –0.25 | –0.10 | –0.15 | –0.15 | –0.15 | –0.15 | 0.00 |
| **SRIC** | | | | | | | | | | – | 0.60 | 0.64 | 0.64 | 0.64 | 0.64 | 0.60 |
| **SumPaths** | | | | | | | | | | | – | 0.96 | 0.96 | 0.96 | 0.96 | 0.73 |
| **MaxPath** | | | | | | | | | | | | – | 1.00 | 1.00 | 1.00 | 0.69 |
| **MaxMin** | | | | | | | | | | | | | – | 1.00 | 1.00 | 0.69 |
| **MultT** | | | | | | | | | | | | | | – | 1.00 | 0.69 |
| **MaxT** | | | | | | | | | | | | | | | – | 0.69 |
| **Simul** | | | | | | | | | | | | | | | | – |

**Table 23.** Goodman, Kruskal γ-coefficient

| | $C^{w\,in-deg}$ | $C^{w\,out-deg}$ | $C^{w\,deg}$ | $C^{w\,deg-diff}$ | $C^{btw}$ | $C^{cl}$ | $C^{PageRank}$ | $C^{eig}$ | $C^{Bonacich}$ | SRIC | SumPaths | MaxPath | MaxMin | MultT | MaxT | Simul |
|---|---|---|---|---|---|---|---|---|---|---|---|---|---|---|---|---|
| $C^{w\,in-deg}$ | – | –0.33 | 1.00 | –0.78 | 1.00 | 0.56 | 1.00 | 1.00 | 1.00 | –0.33 | –1.00 | –1.00 | –1.00 | –1.00 | –1.00 | –0.56 |
| $C^{w\,out-deg}$ | | – | 0.73 | 0.91 | 0.54 | –0.51 | 0.54 | 0.46 | –0.68 | 0.42 | 0.11 | 0.16 | 0.16 | 0.16 | 0.16 | 0.02 |
| $C^{w\,deg}$ | | | – | 0.64 | 0.77 | –0.33 | 0.81 | 0.73 | –0.33 | 0.24 | –0.16 | –0.11 | –0.11 | –0.11 | –0.11 | –0.16 |
| $C^{w\,deg-diff}$ | | | | – | 0.44 | –0.60 | 0.44 | 0.36 | –0.78 | 0.42 | 0.20 | 0.24 | 0.24 | 0.24 | 0.24 | 0.11 |
| $C^{btw}$ | | | | | – | –0.49 | 0.56 | 0.67 | –0.20 | 0.44 | 0.12 | 0.16 | 0.16 | 0.16 | 0.16 | 0.11 |
| $C^{cl}$ | | | | | | – | –0.21 | –0.14 | 0.50 | –0.56 | –0.42 | –0.47 | –0.47 | –0.47 | –0.47 | –0.42 |
| $C^{PageRank}$ | | | | | | | – | 0.62 | –0.39 | 0.02 | –0.40 | –0.35 | –0.35 | –0.35 | –0.35 | –0.40 |
| $C^{eig}$ | | | | | | | | – | –0.09 | 0.14 | –0.23 | –0.23 | –0.23 | –0.23 | –0.23 | –0.23 |
| $C^{Bonacich}$ | | | | | | | | | – | –0.28 | –0.11 | –0.17 | –0.17 | –0.17 | –0.17 | 0.00 |
| **SRIC** | | | | | | | | | | – | 0.60 | 0.64 | 0.64 | 0.64 | 0.64 | 0.60 |
| **SumPaths** | | | | | | | | | | | – | 0.96 | 0.96 | 0.96 | 0.96 | 0.73 |
| **MaxPath** | | | | | | | | | | | | – | 1.00 | 1.00 | 1.00 | 0.69 |
| **MaxMin** | | | | | | | | | | | | | – | 1.00 | 1.00 | 0.69 |
| **MultT** | | | | | | | | | | | | | | – | 1.00 | 0.69 |
| **MaxT** | | | | | | | | | | | | | | | – | 0.69 |
| **Simul** | | | | | | | | | | | | | | | | – |

## 5. Conclusion

Network analysis plays a significant role in many problem areas. When we study different relationships between elements we often need to identify the most powerful participants, or in terms of graph theory, we need to detect key nodes.

We explored two approaches of the influence measure in a network-graph. The key advantage of this approach with comparison to existing methods is that we consider long-distance connections as well as special attributes of nodes (in the form of thresholds) and group influence on nodes. This allows us to detect hidden key nodes: while classical measures detect explicit powerful nodes our methods also detect nodes that influence other nodes in groups or by long-range interactions.

Another important aspect is that our approach also allow us to estimate the intensity of influence of nodes: due to the fact that on one of the stages of the calculation we get node-to-node influences we can not only estimate the level of influence on other nodes for each node but also the level of influence of all nodes on each node. In other words, the methods admit many ways of aggregation which lead to different interpretations of nodes power.

The first approach is based on the analysis of all simple paths between all pairs of nodes in a graph; such methodology allows us to control all channels of influence. As a result, we obtain 5 versions of long-range interaction index (LRIC index). The second approach is based on the idea of simulations: we sequentially consider different influential groups and track the changes in a graph. This allows us to consider all possible scenarios of nodes cooperation.

As our methods are complex for the calculation we suggest some natural limitations as the length of considered paths, the size of groups, etc.

We demonstrate the consistency of our approaches on some numerical examples which confirm that existing measures do not detect hidden nodes but LRIC indices and simulations are able to identify them.

The developed new centrality measures can be successfully applied to many real life processes.

Предложен новый метод оценки влияния агентов в сетевых структурах, который учитывает индивидуальные атрибуты каждой вершины, индивидуальное и групповое влияние вершин, а также интенсивность их взаимодействия. Такой подход помогает выявить как очевидные, так и скрытые центральные элементы, которые не могут быть определены классическими мерами центральностей или другими индексами оценки влияния.

*Алескеров Фуад Тагиевич*, Национальный исследовательский университет «Высшая школа экономики» (НИУ ВШЭ), Международная лаборатория анализа и выбора решений, Институт проблем управления им. В.А. Трапезникова Российской академии наук (ИПУ РАН), Москва; alesk@hse.ru

*Мещерякова Наталья Геннадьевна*, Национальный исследовательский университет «Высшая школа экономики» (НИУ ВШЭ), Международная лаборатория анализа и выбора решений, Институт проблем управления им. В.А. Трапезникова Российской академии наук (ИПУ РАН), Москва; na-tamesc@gmail.com

*Швыдун Сергей Владимирович*, Национальный исследовательский университет «Высшая школа экономики» (НИУ ВШЭ), Международная лаборатория анализа и выбора решений, Институт проблем управления им. В.А. Трапезникова Российской академии наук (ИПУ РАН), Москва; shvydun@hse.ru

Алескеров Фуад Тагиевич, Мещерякова Наталья Геннадьевна,
Швыдун Сергей Владимирович

# Меры центральности в сетях, основанные на характеристиках вершин, дальних взаимодействиях и групповом влиянии

(*на английском языке*)